\documentclass[12pt,a4paper,superscriptaddress,preprint,dvipdfmx]{revtex4}
\usepackage{graphicx}
\usepackage{amssymb,amsmath}
\usepackage{amsthm} 
\usepackage{bm}
\usepackage{color}

\usepackage{ulem} 

\definecolor{forestgreen}{rgb}{0.33,0.61,0.34}

\newtheorem{proposition}{\bf Proposition}[section]

\begin{document}

\title{Dynamical stability of water distribution networks}
\author{Naoki Masuda}
\affiliation{Department of Engineering Mathematics,
Merchant Venturers Building, University of Bristol,
Woodland Road, Clifton, Bristol BS8 1UB, United Kingdom}
\email{naokimas@buffalo.edu}
\author{Fanlin Meng}
\affiliation{Centre for Water Systems, College of Engineering, Mathematics and Physical Sciences, University of Exeter, Exeter, EX4 4QF, United Kingdom}




\begin{abstract}
Water distribution networks are hydraulic infrastructures that aim to meet water demands at their various nodes. Water flows through pipes in the network create nonlinear dynamics on networks. A desirable feature of water distribution networks is high resistance to failures and other shocks given to the system. Such threats would at least transiently change the flow rate in various pipes, potentially undermining the functionality of the whole water distribution system. Here we carry out a linear stability analysis for a nonlinear dynamical system representing the flow rate through pipes that are interconnected through an arbitrary pipe network with reservoirs and consumer nodes. We show that the steady state is always locally stable and develop a method to calculate the eigenvalue that corresponds to the mode that decays the most slowly towards the equilibrium, which we use as an index for resilience of the system. We show that the proposed index is positively correlated with the recovery rate of the pipe network, which was derived from a realistic and industrially popular simulator.
%
%
The present analytical framework is expected to be useful for deploying tools from nonlinear dynamics and network analysis to designing, resilience managements and scenario testings of water distribution networks. 
\end{abstract}

\maketitle

\section{Introduction}

Water distribution systems aim to securely provide drinking water to consumers and for firefighting and hence are key infrastructures in our society.
They are subjected to drastic changes in demands on the daily and seasonal timescales \cite{Butler2006book} and to various threats such as contamination \cite{Rasekh2014EnvirModelSoft} and power outage \cite{Gullick2005JWaterSupplyResTechAqua}. 
For safe and secure water supply, it is important that a water distribution system is well designed and operated against potential failures.

In risk management in water engineering, the risk is estimated to be equal to the product of the probability of failures and the consequence of the failure
\cite{Jun2007JEnvironEng,Giustolisi2010UrbanWaterJ}. In contrast to risk management,
resilience management is a relatively new framework 
to aid us to prepare for potential failures in water distribution systems.
Resilience is a broad concept and its definition varies according to studies in engineering \cite{Hosseini2016ReliabilityEngSystSafety} and in other fields \cite{XueWangYang2018NatHazards}. The concept of resilience is equally wide for water distribution systems; it is construed as the general capacity of a system to resist, absorb, withstand and recover from stresses \cite{Hosseini2016ReliabilityEngSystSafety,Butler2017GlobalChallenges}. 
It is a useful complement, as many failures that happen in real-life are unforeseeable and are not targeted in the traditional risk management due to the low estimated probability.

Water distribution systems are composed of pipes combined with other functional structures
such as reservoirs, pumps and valves. Network disfunctions such as pipe failure are an obvious threat to the functionality of the water distribution system. Network science has accumulated knowledge on how network structure shapes the network's capacity to withhold random failures and intentional attacks \cite{Newman2010book,Barabasi2016book}. Therefore, there has been a surge of interest in water engineering community to deploy network analysis to assess the degree of resilience of water distribution networks.
%
%
However, to the best of our knowledge, most of the existing proposals of resilience measures for water distribution networks are based on the network structure 
\cite{Yazdani2011Chaos,Yazdani2011EnvironModelSoftware,Yazdani2012WaterResourcesRes,ShuangZhangYuan2014PlosOne,Herrera2016WaterResourcesMan,Pundit2016IntJCritInfrastruct,Hwang2017JWaterResourPlannManag,Meng2018WaterRes}, or
steady-state or quasi-steady-state flow rates
\cite{Todini2000UrbanWater,Zhuang2013JHydraulEng,Yannopoulos2013WaterResourManag,Zhao2015MathProbEng,Diao2016WaterRes,Liu2017JWaterResourPlannManag}; as such, the dynamic, transient water flows in the pipe network cannot be represented, which is in fact the key to the resilience performance of a water distribution system. Furthermore, one needs to apply stresses to the system to measure resilience \cite{Meng2018WaterRes}, which cannot be exhaustive of what failure might occur in a system. Therefore, it is necessary to examine transient dynamics to reveal much of resilience features of the system, in addition to static network structure or flow.
In fact, transient dynamics of water flow in pipe networks are not a new issue in water engineering. 
Examination of transient flows is also a practical requirement because live water distribution networks are incessantly undergoing changes due to, for example, routine operational adjustments, breakdowns of their elements and human error. In turn, transient water flows are known to cause various problems in water distribution systems such as pump failure, collapse of vapor cavities and compromised water quality. Motivated by these needs, various methods to simulate and understand transient flows have been developed \cite{Islam1998JHydraulEng,Boulos2006book,Todini2011JHydroinfo,Nault2016JHydraulEng,JungKarney2017UrbanWaterJ}. While useful, these developments are mathematical modelling of the transient water dynamics and their numerical simulation methods. These methods do not themselves provide a measure of resilience.

In the present study, we propose an index of resilience via linear stability analysis of a standard set of dynamical equations modelling the steady-state flow in the given water distribution network. The equations are equivalent to a type of a nonlinear electric circuit system. The proposed resilience index is defined in terms of the dominant eigenvalue of the Jacobian matrix around the steady state. Therefore, the index corresponds to the speed at which the transient dynamics decay towards the steady state in response to a small perturbation in the flow rate.
The index is calculated from the structure of the network, the diameter and length of the individual pipes, and other constraints such as the energy level at the reservoir nodes and water demand at individual consumer nodes. We test the proposed index against different resilience measures that we previously proposed based on realistic numerical simulations of water distribution systems.
MATLAB code for calculating the so-called local stability index is available at Github
(\url{https://github.com/naokimas/LSI_water_distribution_net}).
%

\section{Model}

Consider a connected and undirected network of pipes. By definition, a pipe connects two nodes, where nodes are junctions, reservoirs and households. A dead-end node that is only connected to a single pipe (hence, the node's degree is equal to one) corresponds to a consumer node such as a household. It should be noted that nodes whose degree is larger than one are junction nodes but can also demand water if they are connected to households. Let us denote by $N$ and $M$ the number of nodes and edges, respectively. Each edge has its own diameter and length, assuming a circular shape of the pipe. These properties affect how much water flows through the pipe under a given condition, thus effectively specifying the weight (i.e., capacity) of the edge. However, we consider the network as unweighted network and explicitly model the effect of the pipe diameter and length.

We consider a rigid water column model on pipe networks
under the assumption that flows in a pipe vary slowly in time
\cite{Wylie1983JHydraulEng,Islam1998JHydraulEng,Ghidaoui2005ApplMechRev,Chaudhry2014book}. The dynamical equation representing transient flow through the $m$th pipe ($1\le m\le M$), which connects the $i$th and $j$th nodes, is given by
\begin{equation}
I_m\frac{{\rm d}Q_m}{{\rm d}t} = h_i - h_j - R_m Q_m |Q_m| - u(Q_m).
\label{eq:dyn}
\end{equation}
Nodes $i$ and $j$ depend on edge $m$. However, unless we state otherwise, we use $i$ and $j$ here and in the following text to avoid notational abuse.
In Eq.~\eqref{eq:dyn}, $Q_m$ is the time-dependent flow rate through the $m$th pipe, in the direction from the $i$th node to the $j$th node; $t$ represents the time; $h_i$ represents the total energy at the $i$th node, which depends on the time;
$R_m$ is a constant called the pipe coefficient containing the information on the diameter, length and roughness of pipe $m$; $R_m Q_m |Q_m|$ accounts for the nonlinear head loss across the $m$th pipe; $u(Q_m)$ is the input term which comes from a valve or other structure located on the $m$th pipe; $I_m$ is the pipe inertia constant. The dynamical system given by Eq.~\eqref{eq:dyn} is equivalent to that of a nonlinear resistor-inductor electrical circuit, where $Q_m$ is the current, $h_i$ is the voltage, $I_m$ is the inductance, $R_m$ is a nonlinear resistance and $u(Q_m)$ is other nonlinear elements placed on the $m$th edge.

The pipe coefficient is given by
\begin{equation}
R_m = \frac{8 f_m \ell_m}{g \pi^2 d_m^5},
\label{eq:pipe coefficient}
\end{equation}
where $f_m$ is the dimensionless Darcy-Weisbach friction factor for the $m$th pipe and depends on the flow regime (see below for the formula) and other factors, 
$\ell_m$ is the length of the $m$th pipe, $g$ is the gravitational acceleration and equal to $9.81$ m/s${}^2$ and $d_m$ is the diameter of the $m$th pipe 
\cite{Boulos2006book}. More generally, if the nonlinear head loss term is given by $R_m Q_m 
|Q_m|^{\gamma-1}$, one obtains
$R_m = f_m \ell_m / 2gd_m \mathcal{A}_m^{\gamma}$, where
$\mathcal{A}_m$ is the cross-sectional area of the $m$th pipe, i.e., $\mathcal{A}_m = \pi d_m^2/4$ \cite{Todini1988chapter,Boulos2006book}. The pipe inertia constant is given by $I_m = \ell_m/(g \mathcal{A}_m)$.
%
%

There are various formula to approximate the friction factor, $f_m$, as a function of the pipe diameter, $d_m$, the roughness coefficient of the pipe (which is determined by the pipe's material, age and other factors), Reynolds number and so forth. Because the Reynolds number is a function of the flow rate, $Q_m$, the friction factor also depends on $Q_m$. Our index developed in section~\ref{sec:theory} involves differentiation of the right-hand side of Eq~\eqref{eq:dyn} with respect to $Q_m$, which requires the derivative of $f_m$ with respect to $Q_m$. In fact, many formula for $f_m$ can be only applicable to particular flow regimes; the flow regime is 
specified by the value of the Reynolds number. Because $f_m$ may depend on pipes, different pipes may be in different flow regimes and some pipes may be situated near the boundary of two flow regimes.
Therefore, we select the formula proposed in Ref.~\cite{Bellos2018JHydraulicEng} that is applicable in all flow regimes including laminar and turbulent flows and gives $f_m$ as a continuous function of the Reynolds number (and hence continuous in $Q_m$). The formula is given by
\begin{equation}
f_m = \left(\frac{64}{\mathcal{R}_m}\right)^a \left(0.75 \ln \frac{\mathcal{R}_m}{5.37}\right)^{2(a-1)b}
\left(0.88 \ln \frac{6.82 d_m}{\epsilon}\right)^{2(a-1)(1-b)}.
\label{eq:Bellos}
\end{equation}
In Eq.~\eqref{eq:Bellos}, the Reynolds number for the $m$th pipe is equal to
\begin{equation}
\mathcal{R}_m = \frac{\left|Q_m\right| d_m}{\mathcal{A}_m v},
\end{equation}
where $v = 1.007\times 10^{-6}$ m${}^2/$s is the kinematic viscosity of water, assuming the temperature of 20 degree celsius.
Furthermore, $\epsilon = 2.591 \cdot 10^{-4}$ m is the roughness coefficient of cast iron
\cite{Rossman2000EPANET2} (also see Ref.~\cite{Boulos2006book} for a similar value), and
%
%
%
\begin{align}
a =& \left[1+\left(\frac{\mathcal{R}_m}{2712}\right)^{8.4}\right]^{-1},\\
b =& \left[ 1 + \left( \frac{\epsilon \mathcal{R}_m}{150 d_m}\right)^{1.8}\right]^{-1}.
\end{align}

\section{\label{sec:theory}Local stability index}

In this section, we carry out the linear stability analysis of the steady state of the nonlinear dynamical system given by Eq.~\eqref{eq:dyn}. The system turns out to be always linearly stable, and we propose a local stability index as the magnitude of the eigenvalue corresponding to the most slowly decaying mode. This task is non-trivial because some nodes act as reservoir of water, other nodes act as consumer of water and the friction factor, $f_m$, depends on the dynamical variable, $Q_m$.

\subsection{Removing redundancy from the dynamical equations}

Assume that the total energy at $N_0$ nodes is fixed. The $N_0$ nodes are typically reservoir whose altitude is higher than typical consumer and junction nodes. Without loss of generality, we assume that $h_{N-N_0+1}$, $h_{N-N_0+2}$, $\ldots$, $h_N$ are fixed.
Then, the set of Eq.~\eqref{eq:dyn} has $N-N_0+M$ unknowns, i.e., $h_i$ ($1\le i\le N-N_0$) and $Q_m$ ($1\le m\le M$), whereas Eq.~\eqref{eq:dyn} only provides $M$ differential equations. The remaining constraints come from the Kirchhoff's current law (i.e. conservation of water mass) at each of the $N-N_0$ nodes whose total energy is not fixed. In other words, we obtain
\begin{equation}
\sum_{j=1; j\neq i}^N A_{ij} Q_{ij} = - \tilde{Q}^{\rm ext}_i\quad (i=1, 2, \ldots, N-N_0),
\label{eq:current law}
\end{equation}
where $Q_{ij}$ is the flow rate through pipe $(i, j)$ from the $i$th to $j$th nodes;
$\tilde{Q}^{\rm ext}_i$ is the external demand (i.e., withdrawal) of water at the $i$th node; $A_{ij} (=A_{ji})$ is the element of the adjacency matrix. In other words, $A_{ij}=1$ if there is a pipe between the $i$th and $j$th nodes. Otherwise, $A_{ij} = 0$.


Equations~\eqref{eq:dyn} and \eqref{eq:current law} imply that
there are $N-N_0+M$ unknowns and the same number of equations for solving the steady state and its linear stability.
Now we erase $h_i$ ($1\le i\le N-N_0$) from Eq.~\eqref{eq:dyn} using Eq.~\eqref{eq:current law} to derive a self-contained set of $M$-dimensional dynamical system. It should be noted that we do not have to erase $h_i$ ($N-N_0+1\le i\le N$) from Eq.~\eqref{eq:dyn} because these $h_i$'s are constant. To erase $h_i$ ($1\le i\le N-N_0$), we differentiate Eq.~\eqref{eq:current law} to obtain
\begin{equation}
\sum_{j=1; j\neq i}^N A_{ij} \frac{{\rm d}Q_{ij}}{{\rm d}t} = 0 \quad (i=1, 2, \ldots, N-N_0).
\label{eq:current law/dt}
\end{equation}
By substituting Eq.~\eqref{eq:dyn} in Eq.~\eqref{eq:current law/dt}, one obtains
\begin{equation}
\sum_{j=1; j\neq i}^N A_{ij}
\frac{h_i - h_j - R_{ij} Q_{ij} |Q_{ij}| - u(Q_{ij})}{I_{ij}} = 0\quad (i=1, 2, \ldots, N-N_0),
\label{eq:current law 2}
\end{equation}
where $I_{ij}$ is the inertia constant of pipe $(i, j)$.
Note that Eq.~\eqref{eq:pipe coefficient} implies that $R_{ij} = R_{ji}$.

We define the $M$-dimensional diagonal matrix $D$ by the diagonal elements $D_{mm} = 1/I_m$ ($1\le m\le M$). We also denote the head vector by $\bm h = (h_1, \ldots, h_N)^{\top}$, where $\top$ denotes the transposition. Furthermore, we denote by $\bm q = (q_1, \ldots, q_M)^{\top}$, of which the $m$th element is given by
\begin{equation}
q_m(Q_m) \equiv R_m Q_m |Q_m| + u(Q_m).
\label{eq:def q_m}
\end{equation}

To rewrite Eq.~\eqref{eq:current law 2}, we introduce the incidence matrix, denoted by $C$, which is an $N\times M$ matrix.
By definition, corresponding to each $m$th edge $(i, j)$, the entries of the $m$th column of $C$ are given by $C_{i,m} = 1$, $C_{j,m} = -1$ and $C_{\ell, m} = 0$ ($1\le\ell\le N, \ell\neq i, j$). Although the edges are undirected, we assign $+1$ and $-1$ to $i$ and $j$, respectively, by arbitrarily selecting either node forming the $m$th edge to which $+1$ is assigned, for later convenience. We represent $C$ in the block form given by
\begin{equation}
C = \begin{pmatrix}
C^{(1)}\\ C^{(2)}
\end{pmatrix},
\end{equation}
where the $(N-N_0)\times M$ matrix $C^{(1)}$ is the part of the incidence matrix corresponding to the nodes whose total energy is not fixed, and 
the $N_0\times M$ matrix $C^{(2)}$ is the part of the incidence matrix corresponding to the nodes whose total energy is fixed.

We further denote the $N\times N$ combinatorial Laplacian matrix of the network by
\begin{equation}
L_{ij} = \begin{cases}
\sum_{j^{\prime}=1; j^{\prime}\neq i}^N A_{ij^{\prime}} & (i = j),\\
- A_{ij} & (i\neq j).
\end{cases}
\label{eq:Laplacian}
\end{equation}
Using 
\begin{equation}
L = C C^{\top},
\end{equation}
we obtain the following block format of the Laplacian matrix:
\begin{equation}
L = 
\begin{pmatrix}
L^{(11)} & L^{(12)}\\
L^{(21)} & L^{(22)}
\end{pmatrix}
=
\begin{pmatrix}
C^{(1)} C^{(1)\top} & C^{(1)} C^{(2)\top}\\
C^{(2)} C^{(1)\top} & C^{(2)} C^{(2)\top}
\end{pmatrix},
\end{equation}
where $L^{(11)}$ is the $(N-N_0) \times (N-N_0)$ matrix corresponding to the nodes whose total energy is not fixed, $L^{(12)}$ is an $(N-N_0) \times N_0$ matrix and so forth.

The combinatorial Laplacian matrix for the network with the same set of edges but with the edge weight given by $D_{mm} = 1/I_m$ is called the conductance matrix in electrical circuit theory
\cite{Dorfler2018ProcIeee}. We denote the conductance matrix by $L_{\rm w}$. The conductance matrix is given by 
\begin{equation}
L_{\rm w} = 
\begin{pmatrix}
L_{\rm w}^{(11)} & L_{\rm w}^{(12)}\\
L_{\rm w}^{(21)} & L_{\rm w}^{(22)}
\end{pmatrix}
= C D C^{\top}.
\end{equation}

Then, we can rewrite Eq.~\eqref{eq:current law 2} as
\begin{equation}
\begin{pmatrix}
L_{\rm w}^{(11)} & L_{\rm w}^{(12)}
\end{pmatrix}
\bm h - C^{(1)} D \bm q = 0.
\label{eq:current law 2 succinct}
\end{equation}
With the notation
\begin{equation}
\bm h = \begin{pmatrix}
\bm{h^{(1)}}\\
\bm{h^{(2)}}
\end{pmatrix},
\end{equation}
where $\bm{h^{(1)}} = (h_1, \ldots, h_{N-N_0})^{\top}$ and $\bm{h^{(2)}} = (h_{N-N_0+1}, \ldots, h_N)^{\top}$,
Eq.~\eqref{eq:current law 2 succinct} is equivalent to
\begin{equation}
C^{(1)}D C^{(1)\top} \bm{h^{(1)}} = - C^{(1)} D C^{(2)\top} \bm{h^{(2)}} + C^{(1)} D \bm q.
\end{equation}
Matrix $C^{(1)}D C^{(1)\top}$ is so-called the loopy Laplacian matrix
\cite{Dorfler2018ProcIeee}. Because its all eigenvalues are positive if the network is a connected network, which we have assumed, $C^{(1)}D C^{(1)\top}$ has the inverse and one obtains
\begin{equation}
\bm{h^{(1)}} = 
\left(C^{(1)}D C^{(1)\top}\right)^{-1}\left( - C^{(1)} D C^{(2)\top} \bm{h^{(2)}} + C^{(1)} D \bm q\right).
\label{eq:h^(1) by h^(2)}
\end{equation}
This procedure is essentially the same as the network Kron reduction \cite{Dorfler2018ProcIeee}.

By substituting Eq.~\eqref{eq:h^(1) by h^(2)} in Eq.~\eqref{eq:dyn}, one obtains
\begin{equation}
\frac{{\rm d}\bm Q}{{\rm d}t} = D C^{\top}
\begin{pmatrix}
\left(C^{(1)}D C^{(1)\top}\right)^{-1} \left( - C^{(1)} D C^{(2)\top} \bm{h^{(2)}} + C^{(1)} D \bm q(\bm Q)\right)\\
\bm{h^{(2)}}
\end{pmatrix}
- D \bm q(\bm Q),
\label{eq:dyn 2}
\end{equation}
where $\bm Q \equiv (Q_1, \ldots, Q_M)^{\top}$. Equation~\eqref{eq:dyn 2} is an $M$-dimensional dynamical system. However, it should be noted that the dynamics are confined on a $(M-N+N_0)$-dimensional hyperplane specified by Eq.~\eqref{eq:current law}, respecting the Kirchhoff's current law.


\subsection{Local stability analysis}

In this section, we derive the eigenequation that determines the conventional local linear stability of the steady state of the nonlinear dynamics given by Eq.~\eqref{eq:dyn}. 
We use the combination of Eqs.~\eqref{eq:dyn} and \eqref{eq:current law} for solving the steady state and
Eq.~\eqref{eq:dyn 2} to perform the local stability analysis around the obtained steady state.

The steady state is given by setting the left-hand side of Eq.~\eqref{eq:dyn} to zero. By combining the resulting $M$ constraints with the $N-N_0$ constraints provided by Eq.~\eqref{eq:current law}, one can obtain the steady state, which consists of the $N-N_0+M$ unknowns, i.e., $h_i$ ($1\le i\le N-N_0$) and $Q_m$ ($1\le m\le M$).

To set up the Newton-Raphson iteration scheme (e.g. \cite{Boulos2006book}), we rewrite 
Eq.~\eqref{eq:current law} as
\begin{equation}
F_i^{\rm node} (\bm Q) \equiv \sum_{j=1; j\neq i}^N A_{ij} Q_{ij} + \tilde{Q}^{\rm ext}_i = 0
\quad (i=1, 2, \ldots, N-N_0).
\label{eq:current law NP}
\end{equation}
We set the left-hand side of Eq.~\eqref{eq:dyn} to 0 to obtain
\begin{equation}
F_m^{\rm pipe}(\bm{h^{(1)}}, \bm Q) \equiv h_{i_m} - h_{j_m} - R_m Q_m |Q_m| - u(Q_m) = 0,
\label{eq:dyn NP}
\end{equation}
where $i_m$ and $j_m$ are the two nodes connected by the $m$th pipe; we use $i_m$ and $j_m$ instead of $i$ and $j$ to avoid possible notational conflict with Eq.~\eqref{eq:current law NP}.
The Jacobian of the right-hand side of Eqs.~\eqref{eq:current law NP} and \eqref{eq:dyn NP} with $N+M$ unknowns, $h_1$, $\ldots$, $h_{N-N_0}$, $Q_1$, $\ldots$, $Q_M$, is given by
\begin{equation}
J = \begin{pmatrix}
O & C^{(1)}\\
\left(C^{(1)}\right)^{\top} & - \text{diag}(q_m^{\prime}(Q_m))
\end{pmatrix},
\end{equation}
where $\text{diag}(q_m^{\prime}(Q_m))$ 
is the diagonal matrix whose diagonal elements are given by
\begin{equation}
q_m^{\prime}(Q_m) = 2 R_m \text{sgn}(Q_m)Q_m + u^{\prime}(Q_m).
\end{equation}
The update equation for the Newton-Raphson method is given by
\begin{equation}
\begin{pmatrix}
\bm h^{(1)(n+1)}\\
\bm Q^{(n+1)}
\end{pmatrix}
=
\begin{pmatrix}
\bm h^{(1)(n)}\\
\bm Q^{(n)}
\end{pmatrix}
- J^{-1}
\begin{pmatrix}
\bm F_1^{\rm node}(\bm Q^{(n)})\\
\vdots\\
\bm F_{N-N_0}^{\rm node}(\bm Q^{(n)})\\
\bm F_1^{\rm pipe}(\bm h^{(1)(n)}, \bm Q^{(n)})\\
\vdots\\
\bm F_M^{\rm pipe}(\bm h^{(1)(n)}, \bm Q^{(n)})
\end{pmatrix}.
\end{equation}

Denote the obtained steady state by $(h_1^*, \ldots, h_{N-N_0}^*, Q_1^*, \ldots, Q_M^*)$.
We now carry out the local stability analysis around this steady state.
The Jacobian matrix of the right-hand side of Eq.~\eqref{eq:dyn 2}, denoted by
$J^{\rm dyn} = (J^{\rm dyn}_{m m^{\prime}})$, is given by
\begin{equation}
J^{\rm dyn} = \left[ D C^{(1)\top} \left(C^{(1)}DC^{(1)\top}\right)^{-1} C^{(1)} D -  D \right]
\text{diag}\left(q_m^{\prime}(Q_m^*) \right).
\label{eq:J^dyn}
\end{equation}

\bigskip

\noindent
\begin{proposition}
Matrix $B \equiv D C^{(1)\top} \left(C^{(1)}DC^{(1)\top}\right)^{-1} C^{(1)} D -  D$ has $(N-N_0)$-fold zero eigenvalues and $M-N+N_0$ negative eigenvalues.
\end{proposition}

\bigskip

\begin{proof}
A direct substitution verifies
\begin{equation}
B (C^{(1)})^{\top} = 0.
\end{equation}
Therefore, $B$ has $(N-N_0)$-fold zero eigenvalues, and the corresponding $N-N_0$ right eigenvectors are the columns of $(C^{(1)})^{\top}$.

Because $\left(C^{(1)}DC^{(1)\top}\right)^{-1}$ is an $(N-N_0) \times (N-N_0)$ matrix, its rank is at most $N-N_0$. Therefore, the $M\times M$ matrix $B^{\prime} \equiv C^{(1)\top} \left(C^{(1)}DC^{(1)\top}\right)^{-1} C^{(1)} D$ has at most rank $N-N_0$ and therefore has at least $M-N+N_0$ zero eigenvalues. 

We denote the eigenvalues of a symmetric matrix $X$ by $\lambda_1(X) \le \cdots \le \lambda_M(X)$.
Using the fact that $D$ is positive definite because all its diagonal elements are positive (i.e., $= 1/I_m$), the 
Weyl's theorem on eigenvalues applied to the sum of $B^{\prime}$ and $-D$, which is equal to $B$, implies that 
\begin{equation}
\lambda_m(B) < \lambda_m(B^{\prime}).
\label{eq:B<B'}
\end{equation}
Note that $\lambda_{m_0}(B^{\prime}) = \lambda_{m_0+1}(B^{\prime}) = \cdots = \lambda_{m_0+M-N+N_0-1}(B^{\prime}) = 0$ for an $m_0$ ($1\le m_0\le N-N_0+1$). Therefore, Eq.~\eqref{eq:B<B'} yields
$\lambda_{m_0}(B), \ldots, \lambda_{m_0+M-N+N_0-1}(B) < 0$. Because this relationship must be consistent with the fact that $B$ has $(N-N_0)$-fold zero eigenvalues, we obtain $m_0=1$ and
$\lambda_1(B), \ldots, \lambda_{M-N+N_0}(B) < 0, \lambda_{M-N+N_0+1}(B) = \cdots = \lambda_M(B) = 0$.
\end{proof}

Let us assume for any pipe $m$ that $q_m^{\prime}(Q_m^*) > 0$, which means that the head loss owing to friction along the pipe increases as the steady-state flow rate, $Q_m^*$, increases. We verified $q_m^{\prime}(Q_m^*) > 0$ held true in all the numerical simulations carried out in the next section. Then, $J^{\rm dyn}$ has $M-N+N_0$ negative eigenvalues and $N-N_0$ zero eigenvalues as matrix $B$ does (Appendix~A).
%

Therefore, the dynamics given by Eq.~\eqref{eq:dyn 2} are locally neutrally stable in the subspace spanned by the column vectors of $\left[\text{diag}\left(q_m^{\prime}(Q_m^*) \right)\right]^{-1} (C^{(1)})^{\top}$. However, perturbations along these directions are infeasible because such a perturbation is not on the hyperplane defined by Eq.~\eqref{eq:current law} on which the dynamics are constrained.
Note that the normal vectors of the hyperplane are given by the row vectors of $C^{(1)}$, i.e., column vectors of $(C^{(1)})^{\top}$.
Therefore, in the local stability analysis, we should ignore these $N-N_0$ zero eigenvalues and inspect the other $M-N+N_0$ eigenvalues. With this consideration, we conclude that the steady state constrained by the Kirchhoff's current law (i.e., Eq.~\eqref{eq:current law}) is always locally stable.

%

To define a local stability index,
we calculate the largest eigenvalue of $J^{\rm dyn}$ except the $(N-N_0)$-fold zero eigenvalue. Note that all eigenvalues of $J^{\rm dyn}$ are real although $J^{\rm dyn}$ is not a symmetric matrix in general
(Appendix~A).
%
%
We consider the 
the largest non-zero eigenvalue of $J^{\rm dyn}$ (which is negative) and refer to its absolute value as the local stability index, denoted by $\rho$. A large $\rho$ value implies that dynamics perturbed by a small amount from the steady state would rapidly return to the steady state. Mathematically, the perturbations are given to the flow rate for some pipes. However, we expect that $\rho$ serves as an index that can characterise dynamical stability of water distribution networks in wider contexts.

\section{Numerical results}

In this section, we calculate the local stability index for various networks and examine its relevance to stability and resilience in simulated dynamics of water flow.

We use the synthesized networks used in our previous study \cite{Meng2018WaterRes}. As supplementary materials of Ref.~\cite{Meng2018WaterRes}, the full network structure and properties of system components (e.g., pipes, nodal demands, elevations) were made open to the public in the EPANET2 format, except for the one network. For completeness, we have made the complete data for all the networks and their system components used in the present study available in the MATLAB format 
at Github, alongside the code for calculating the local stability index.

Of the 85 networks used in our previous study \cite{Meng2018WaterRes}, five networks include pumps, which would pin the flow rate to a constant value. Because our formalism does not directly cover the case of the pinned flow rate, we exclude these five networks. Of the remaining 80 networks, 10 networks have $N=102$ nodes, $N_0=2$ reservoir nodes among the $N$ nodes and $M=110$ pipes; 11 networks have $(N, N_0, M) = (102, 2, 130)$; 10 networks have $(N, N_0, M) = (204, 4, 223)$; 10 networks have $(N, N_0, M) = (204, 4, 263)$; 9 networks have $(N, N_0, M) = (306, 6, 335)$; 9 networks have $(N, N_0, M) = (306, 6, 395)$; 5 networks have $(N, N_0, M) = (404, 4, 443)$; 5 networks have $(N, N_0, M) = (404, 4, 523)$; 5 networks have $(N, N_0, M) = (406, 6, 525)$; 5 networks have $(N, N_0, M) = (408, 8, 447)$; 1 network has $(N, N_0, M) = (506, 6, 554)$.
We calculated $\rho$ for each of the 80 networks.

The diameter of a majority of pipes was equal to 400 mm. There were also a few larger values of the diameter (e.g., 900 mm). Such wider pipes were incident (i.e., directly connected) to a reservoir.
The diameter of the pipes was automatically generated by the software \cite{Decorte2014WaterResourcesMan}.

We set the total energy of all the $N_0$ reservoirs in each network to 65 m \cite{Decorte2014WaterResourcesMan}. In addition to these $N_0$ nodes, some nodes in each network had zero water demand (i.e., $\tilde{Q}_i^{\rm ext} = 0$) because they are pure junctions connecting different pipes.
We set $u(Q_m)$ for each pipe to zero because the data are from the 80 networks that do not have valves or pumps.



In our previous study, we numerically simulated dynamics of water flow in these networks using
software EPANET2 \cite{Mugume2015WaterRes,Diao2016WaterRes} and measured six strain indices to characterise resilience of water distribution systems. These indices are as follows
(see \cite{Meng2018WaterRes} for fuller definitions). First, the time to strain is defined as the time between the application of stress and the start of service failure, which is when the level of service at a node drops below a predefined threshold. Second, the failure duration is defined as the time needed for the system to recover to normal performance since the service failure triggered by the application of stress. Third, the failure magnitude is defined as the most severe drop at a node in terms of the system service performance as a result of the administered stress, averaged over all the nodes. Fourth, the failure rate is defined as the failure magnitude divided by the time between the start of the failure and the occurrence of the worst system performance. Fifth, the recovery rate is defined as the failure magnitude divided by the time between the worst system performance and the return to the performance threshold value used in the definition of the time to strain. Sixth, the severity is defined as the threshold minus the system performance (which is positive during the system failure) integrated with respect to time over the period of system failure.

\begin{figure}[t]
\begin{center}
\includegraphics[scale=0.33]{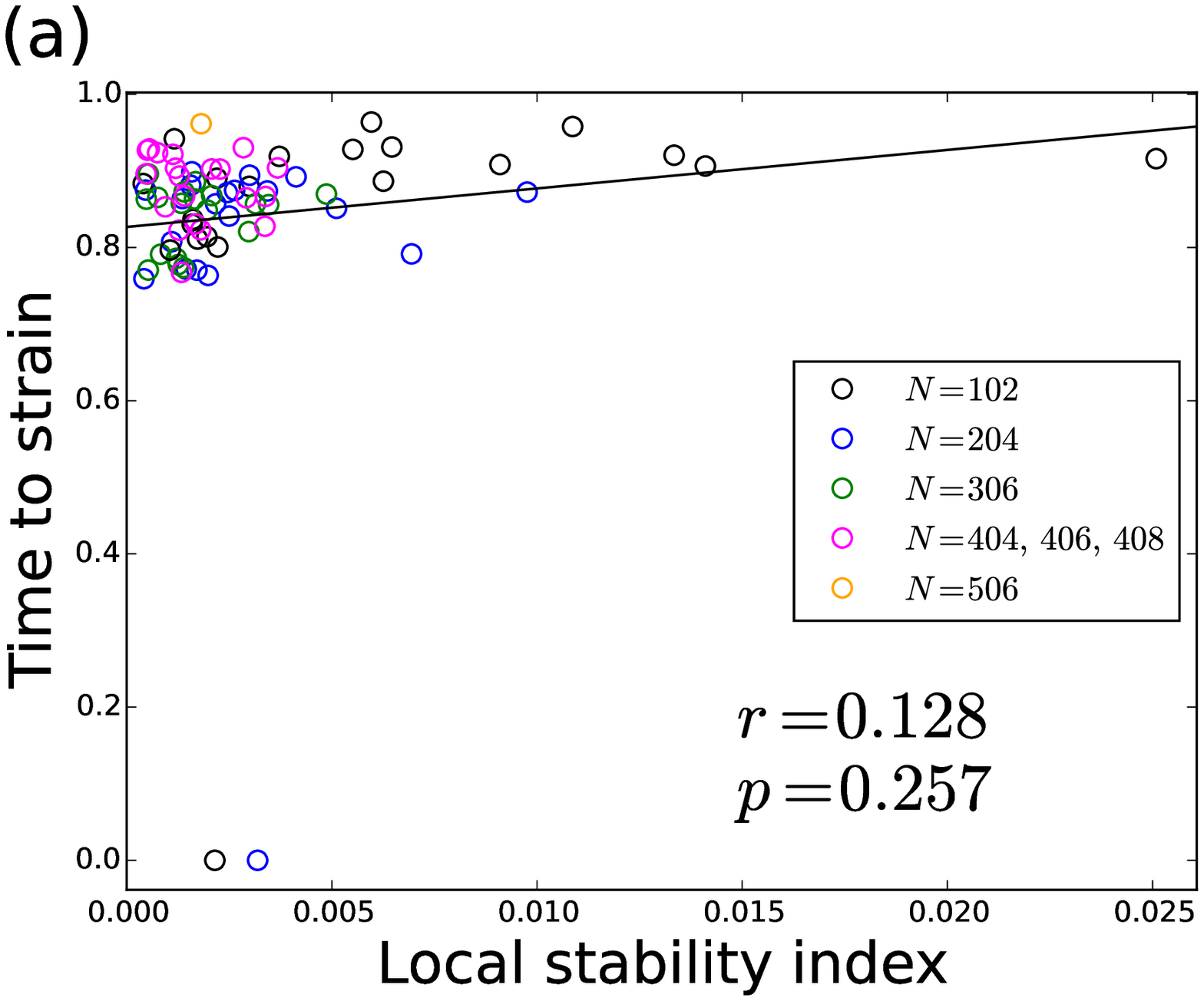}
\includegraphics[scale=0.33]{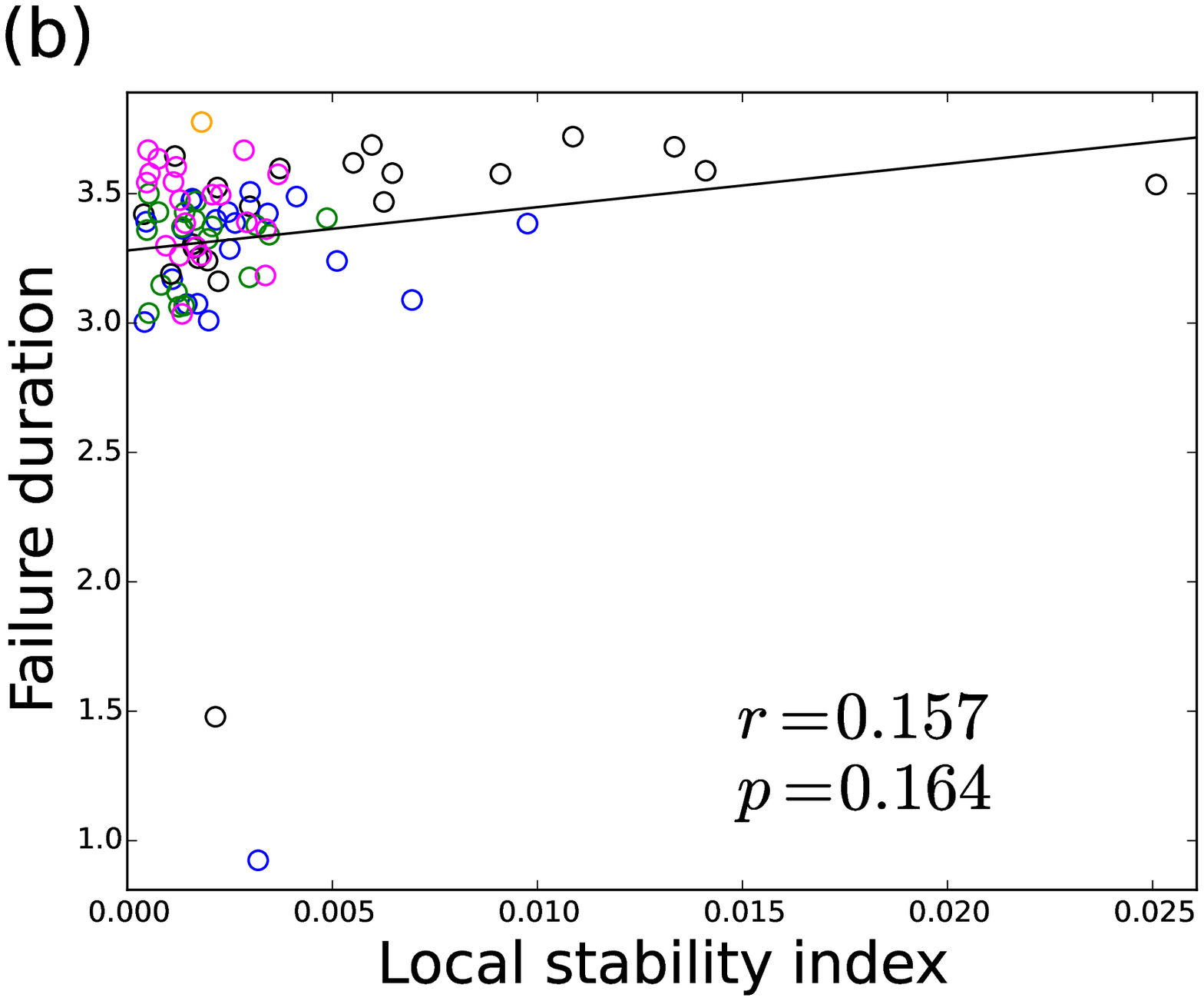}
\includegraphics[scale=0.33]{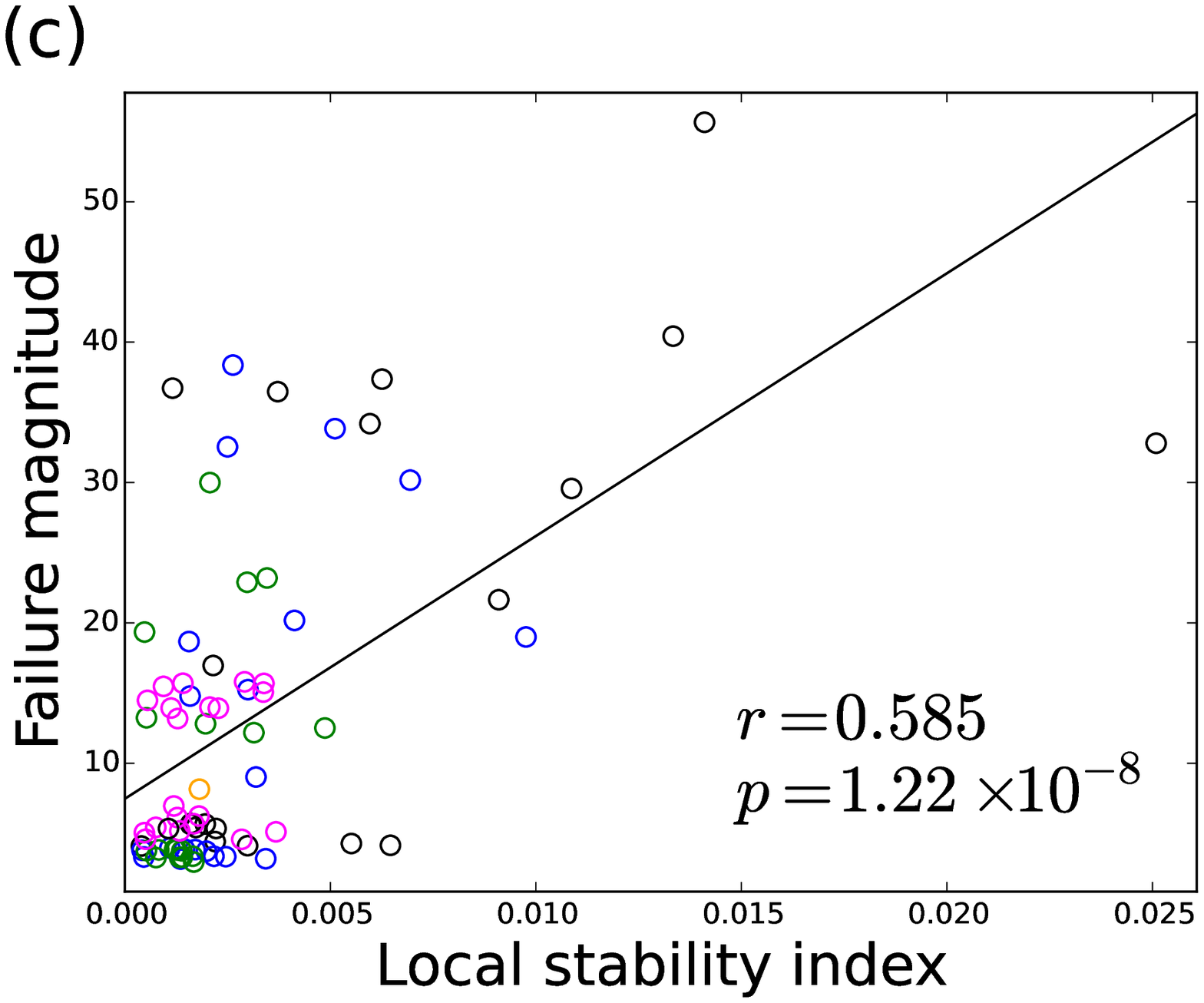}
\includegraphics[scale=0.33]{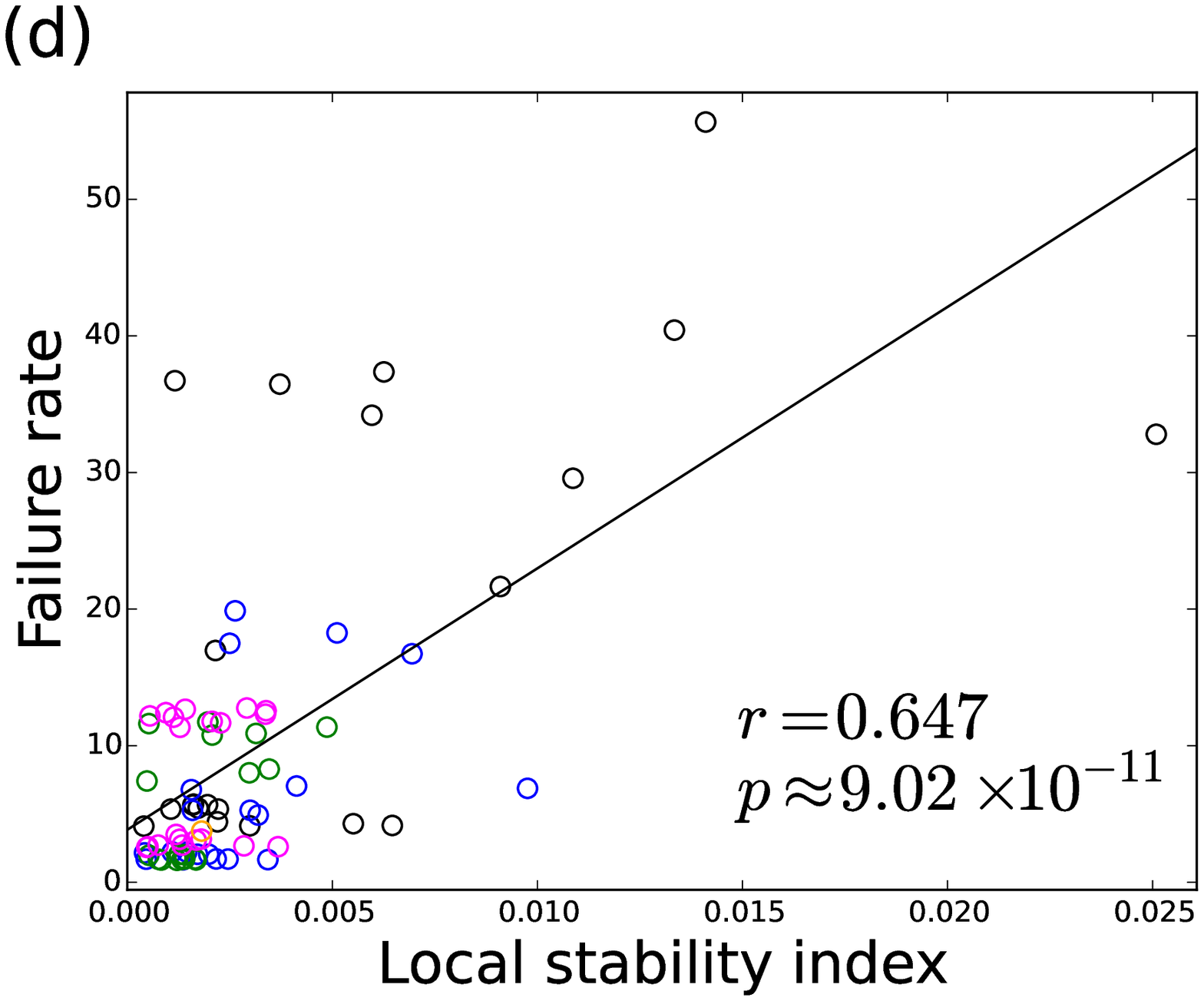}
\includegraphics[scale=0.33]{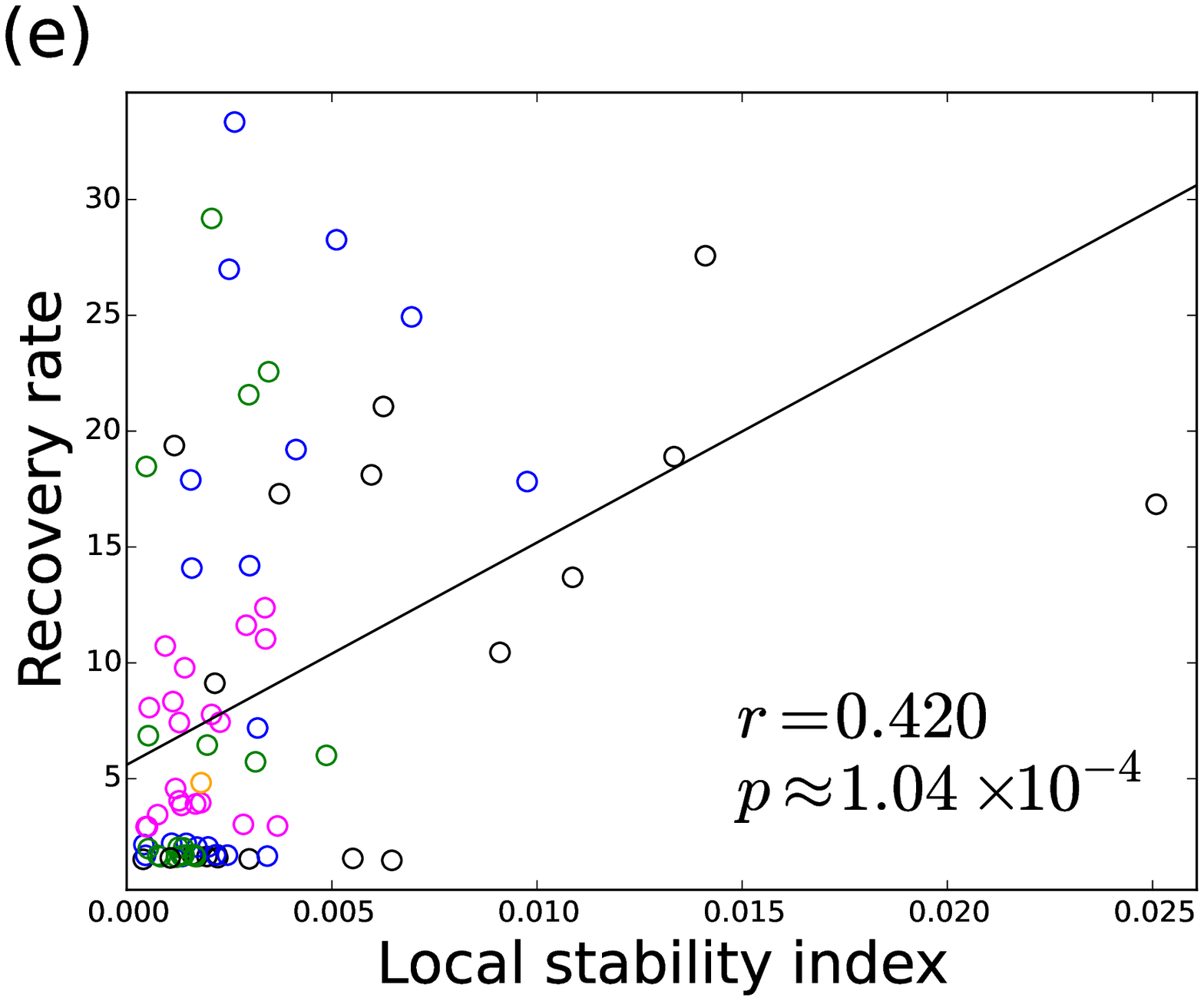}
\includegraphics[scale=0.33]{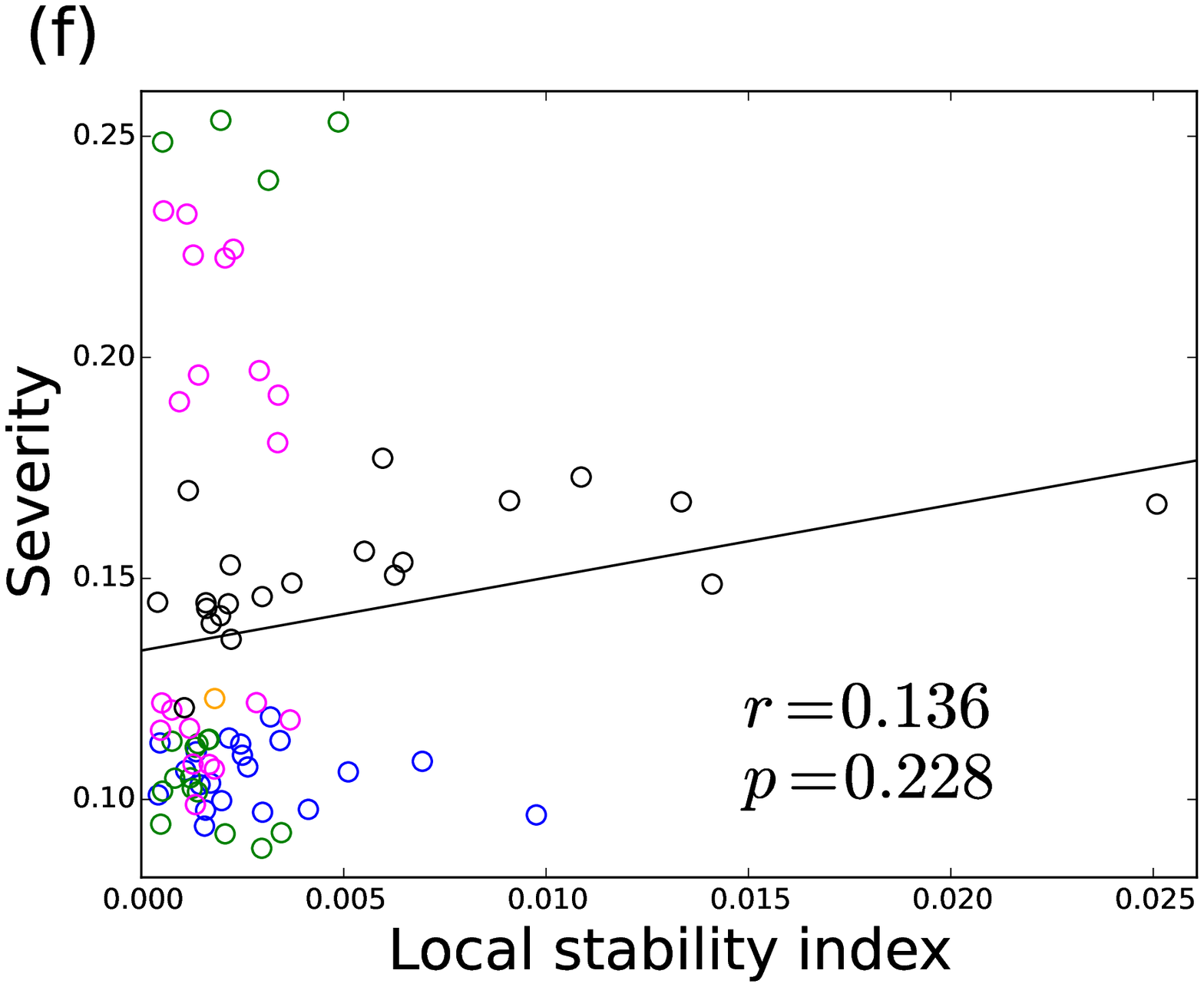}
\caption{Relationship between each of the strain indices and the local stability index. (a) 
Time to strain. (b) Failure duration. (c) Failure magnitude. (d) Failure rate. (e) Recovery rate. (f) Severity. A circle represents a network and its colour represents the number of nodes. The lines represent the linear regression. The Pearson correlation coefficient and its $p$-value are denoted by $r$ and $p$, respectively.}
\label{fig:vs local stability index}
\end{center}
\end{figure}

The relationship between each of the six strain indices and the local stability index, $\rho$, is shown in
Fig.~\ref{fig:vs local stability index}. In the figure, each circle represents one of the 80 networks,
$r$ represents the Pearson correlation coefficient and $p$ represents the $p$-value for the Pearson correlation coefficient. We found a positive correlation between $\rho$ and the recovery rate (Fig.~\ref{fig:vs local stability index}(e)). This result is intuitive
because the recovery rate quantifies how the system returns to the normal performance after being perturbed by external stress, which agrees with the concept of linear stability when the given external stress is infinitesimally small. 
The obtained correlation is only moderate (i.e., $r=0.420$, $p\approx 1.04\times 10^{-4}$) probably due to various nonlinearity of the system, which our linear stability analysis does not aim to explain. However, we consider that $\rho$ captures a strain index derived from EPANET2, which is an involved numerical simulation package, reasonably well.

Figure~\ref{fig:vs local stability index} also indicates that
$\rho$ is strongly correlated with the failure magnitude (Fig.~\ref{fig:vs local stability index}(c)) and the failure rate (Fig.~\ref{fig:vs local stability index}(d)).
We do not have an active interpretation of this result.
In fact, it apparently sounds contradictory that a large $\rho$ value, which predicts that the
network is resilient, is associated with a large failure magnitude and failure rate.
However, the failure magnitude and failure rate quantify the magnitude of failure response induced by external stress, which does not have to do with our local stability analysis even in the limit of infinitesimally small external stress.
Therefore, we conclude that the failure magnitude and the failure rate characterise different aspects of the system's resilience from what the recovery rate and $\rho$ do.
This view is consistent with our additional observation that,
within our previous numerical results \cite{Meng2018WaterRes}, there is positive correlation between the failure magnitude and the recovery rate 
($r=0.917$, $p<10^{-32}$)
and between the failure rate and the recovery rate 
($r=0.677$, $p\approx 5.60\times 10^{-12}$).

In our previous study, we measured eight structural properties of the water distribution network and examined the association between each of them and each of the six strain indices \cite{Meng2018WaterRes}.
Here we measured the Pearson correlation between each of these structural properties, whose definitions are in 
Appendix~B, and the recovery rate. We found that none of the eight structural properties of the network was as strongly correlated with the recovery rate as $\rho$ was (link density: $r = 0.145$, $p = 0.204$; algebraic connectivity: $r = 0.114$, $p = 0.316$;
diameter: $r = -0.143$, $p = 0.207$;
average path length: $r = -0.204$, $p = 0.0717$;
central point dominance: $r = -0.0515$, $p = 0.652$;
heterogeneity: $r = 0.335$, $p = 0.0026$;
spectral gap: $r = -0.0509$, $p = 0.656$;
clustering coefficient: $r = 0.198$, $p = 0.0798$).
Therefore, the local stability index is better at capturing the recovery rate than these structural properties. This is probably because these structural properties do not directly quantify the speed at which the water distribution network responds to a perturbation to the steady state.

\section{Discussion}

We carried out a local stability analysis of water flow in pipe networks including reservoir nodes and consumer nodes. The steady state was shown to be always linearly stable. We used the eigenvalue corresponding to the slowest relaxation mode as the local stability index. The proposed index was moderately correlated with the recovery rate in response to external stress, which was numerically obtained from an involved simulator commonly used in water engineering research community.

We used a particular formula for the friction factor 
\cite{Bellos2018JHydraulicEng}. However, the present framework accepts other formulae for the friction factor. Because the differentiation of the friction factor with respect to the Reynolds number is used in calculating the local stability index and the value of the friction factor depends on the pipe in general, one should use a formula that is continuous and covers a wide range of the Reynolds number and hence different flow schemes. Some possible choices of the formula different from the one used in the present paper are found in the literature \cite{Swamee1993JTransportEng,Cheng2008JHydraulicEng,Brkic2018ApplSci}. It is straightforward to extend our local stability index to the case of different functional forms of the friction factor.
 
In 1972, May proposed to examine the eigenvalue spectra of interaction networks to characterise dynamical stability of complex systems, in particular ecological systems \cite{May1972Nature-stability}. This is
a landmark example of local stability analysis.
While the original analysis and many studies that ensued were for random networks \cite{May1972Nature-stability}, the understanding has been extended to the case of interaction networks with general network structure \cite{Okuyama2008EcolLett,Allesina2012Nature}. This family of method is similar to the one proposed in the present study. Main differences are that we have started from a particular set of hydraulic equations modelling water flow in pipe networks and that we have found that our system is always locally stable and hence used the relaxation mode to quantify the extent of the local stability of the system.

The proposed local stability index was significantly correlated with the recovery rate, which was obtained from numerical simulations in which the applied shock was not necessarily small. However,
because the present analysis is a local stability analysis, it does not tell us the behaviour of the system when a shock injected to the steady state is not small. In water engineering applications, the magnitude of perturbation is not necessarily small \cite{Gullick2005JWaterSupplyResTechAqua}. A recent study in network science has developed a formalism that reduces a class of dynamical systems on interaction networks
into a one-dimensional effective dynamical system and assesses the resilience of the dynamical system against perturbation that is not necessarily small \cite{GaoBarzelBarabasi2016Nature}. 
Therefore, adapting their resilience formalism to the case of transient and steady-state dynamics of water flow in pipe networks is of a practical relevance. In Eq.~\eqref{eq:dyn}, there are usually only a fraction of nodes whose energy value is fixed (i.e., reservoirs) and different nodes have different water demands. In this sense, the flow dynamics on pipe networks are heterogeneous in addition to being heterogeneous in the network structure and edge weight. Other structures such as valves would add more complexity and heterogeneity to the system. For such a heterogeneous system, it is unclear whether one can transform Eq.~\eqref{eq:dyn} into a form that is compatible with the resilience function formalism \cite{GaoBarzelBarabasi2016Nature,Tu2017PhysRevE}. This issue is left as future work.

\enlargethispage{20pt}


\section*{Data accessibility}

The data on the network structure and system components are available on Github.

\section*{Competing interests}
The authors declare that they have no competing interests.

\section*{Authors' contributions}

NM and FM conceived of and designed the study. FM provided the data on the network structure and system components. NM carried out the theoretical analysis and performed the data analysis.  NM and FM discussed the results and drafted the manuscript. All authors read
and approved the manuscript.

\section*{Funding}
N.M. and F.M. acknowledge the support provided through 
BRIM (Building Resilience into Risk Management) project, EPSRC (EP/N010329/1).


\section*{\label{sec:stability of J^dyn}Appendix A: Eigenvalues of $J^{\rm dyn}$}

Here we provide an elementary proof that $J^{\rm dyn}$ has $M-N+N_0$ zero eigenvalues and $N-N_0$ negative eigenvalues when $q_m^{\prime}(Q_m^*) > 0$, $1\le m\le M$. 

Let us introduce a short-hand notation $\overline{D} \equiv \text{diag}\left(q_m^{\prime}(Q_m^*) \right)$. Then, one obtains $J^{\rm dyn} = B \overline{D}$. Because $J^{\rm dyn}$ is similar to the symmetric matrix $\overline{D}^{1/2} B \overline{D}^{1/2}$, the min-max theorem
%
%
for the $k$th smallest eigenvalue implies that
\begin{align}
\lambda_k (J^{\rm dyn}) = \lambda_k \left(\overline{D}^{1/2} B \overline{D}^{1/2}\right)
=& \min_{\substack{S\subset \mathbb{R}^M\\ \dim (S) = k}}
\left[ \max_{x\in S\backslash \{0\}} \frac{\left(\overline{D}^{1/2} B \overline{D}^{1/2}x, x\right)}{(x, x)}\right]\notag\\
=& \min_{\substack{S\subset \mathbb{R}^M\\ \dim (S) = k}}
\left[ \max_{x\in S\backslash \{0\}} \frac{\left(\overline{D}^{1/2} B \overline{D}^{1/2}x, x\right)}{\left(\overline{D}^{1/2}x, \overline{D}^{1/2}x\right)} 
\cdot \frac{\left(\overline{D}x, x\right)}{(x, x)}\right]\notag\\
\le & \min_{\substack{S\subset \mathbb{R}^M\\ \dim (S) = k}}
\left[ \max_{x\in S\backslash \{0\}} \frac{\left(\overline{D}^{1/2} B \overline{D}^{1/2}x, x\right)}{\left(\overline{D}^{1/2}x, \overline{D}^{1/2}x\right)} 
\cdot
\max_{x^{\prime}\in \mathbb{R}^M\backslash \{0\}}
\frac{\left(\overline{D}x^{\prime}, x^{\prime}\right)}{(x^{\prime}, x^{\prime})}\right]\notag\\
=& \lambda_k(B) \lambda_M\left(\overline{D}\right).
\end{align}
Similarly, one obtains
\begin{align}
\lambda_k (J^{\rm dyn})
=& \min_{\substack{S\subset \mathbb{R}^M\\ \dim (S) = k}}
\left[ \max_{x\in S\backslash \{0\}} \frac{\left(\overline{D}^{1/2} B \overline{D}^{1/2}x, x\right)}{\left(\overline{D}^{1/2}x, \overline{D}^{1/2}x\right)} 
\cdot \frac{\left(\overline{D}x, x\right)}{(x, x)}\right]\notag\\
\ge & \min_{\substack{S\subset \mathbb{R}^M\\ \dim (S) = k}}
\left[ \max_{x\in S\backslash \{0\}} \frac{\left(\overline{D}^{1/2} B \overline{D}^{1/2}x, x\right)}{\left(\overline{D}^{1/2}x, \overline{D}^{1/2}x\right)} 
\cdot
\min_{x^{\prime}\in \mathbb{R}^M\backslash \{0\}}
\frac{\left(\overline{D}x^{\prime}, x^{\prime}\right)}{(x^{\prime}, x^{\prime})}\right]\notag\\
=& \lambda_k(B) \lambda_1\left(\overline{D}\right).
\end{align}
By combining $\lambda_1\left(\overline{D}\right) > 0$, $\lambda_M\left(\overline{D}\right) > 0$ and
$\lambda_1(B), \ldots, \lambda_{M-N+N_0} < 0, \lambda_{M-N+N_0+1} = \cdots = \lambda_M = 0$, one obtains
$\lambda_1(J^{\rm dyn}), \ldots, \lambda_{M-N+N_0}(J^{\rm dyn}) < 0, \lambda_{M-N+N_0+1}(J^{\rm dyn}) = \cdots = \lambda_M(J^{\rm dyn}) = 0$.

\section*{Appendix B: Structural properties of networks}

The eight structural properties of the network that we measured in our previous study \cite{Meng2018WaterRes} and used in the present article for comparison purposes are as follows.

The link density is given by $2M/\left[ N(N-1)\right]$. The algebraic connectivity is the smallest positive eigenvalue of the Laplacian matrix, $L$, of the undirected and unweighted network. The version of the clustering coefficient used in 
\cite{Meng2018WaterRes} is
given by $3\times (\text{number of triangles in the network}) / (\text{number of connected triples of nodes in the network})$. The average path length is equal to the shortest path length between two nodes, which is averaged over all the possible $N(N-1)/2$ node pairs. The central point dominance is given by 
$\sum_{i=1}^N (b_{\max}-b_i)/(N-1)$, where $b_i$ is the betweenness centrality of the $i$th node and $b_{\max} = \max_{j=1, \ldots, N} b_j$. The heterogeneity is that in the degree and given by
$\sum_{i=1}^N \left( k_i - \langle k\rangle\right)^2 / \langle k\rangle$, where $k_i$ is the degree of the $i$th node and $\langle k\rangle = \sum_{i=1}^N k_i/N$ is the average degree of the node. The version of the spectral gap used in \cite{Meng2018WaterRes} is given by the difference between the two largest eigenvalues of $L$.
The modularity is a quantity representing the quality of community structure detected in the network and approximate modularity maximisation was carried out using Newman's algorithm
\cite{Newman2004PhysRevE-fast}.



\newpage
\clearpage

\end{document}